\documentclass[prl,aps,showpacs, twocolumn]{revtex4}

\usepackage{graphicx}
\usepackage{epsfig}

\begin{document}
\title{ Measurement of interaction energy near a Feshbach resonance in
a $^6$Li Fermi gas} \author{T.\,Bourdel$^1$, J.\,Cubizolles$^1$,
L.\,Khaykovich$^1$, K.\,M.\,F.\,Magalh\~aes$^1$,
S.\,J.\,J.\,M.\,F.\,Kokkelmans$^1$, G.\,V.\,Shlyapnikov$^{1,2,3}$, and
C.\,Salomon$^1$} \affiliation{{$^1$Laboratoire Kastler Brossel, Ecole
Normale Sup\'erieure, 24 rue Lhomond, 75231 Paris 05,
France}\\{$^2$FOM Institute AMOLF, Kruislaan 407, 1098 SJ
Amsterdam, The Netherlands}\\{$^3$Russian Research Center
Kurchatov Institute, Kurchatov Square, 123182 Moscow, Russia}}
\date{\today}
\begin{abstract}
We investigate the strongly interacting regime in an optically trapped
$^6$Li Fermi mixture near a Feshbach resonance. The resonance is found
at $800(40)\,$G in good agreement with theory. Anisotropic expansion
of the gas is interpreted by collisional hydrodynamics.  We observe an
unexpected and large shift ($80\,$G) between the resonance peak and
both the maximum of atom loss and the change of sign of the
interaction energy.
\end{abstract}
\maketitle 
The achievement of Bose-Einstein condensation (BEC) in
dilute atomic gases \cite{Fermi99} has naturally triggered research on
cooling of Fermi gases to quantum degeneracy. Several groups have now
reached Fermi degeneracy in $^{40}$K and $^6$Li with temperatures down
to about $0.1$ to $0.2$ of the Fermi temperature $T_\mathrm{F}$
\cite{DeMarco99,Schreck01,Truscott01,Dieckmann02,Modugno02,OHara02}.
One of the major goals of this research is to observe the transition
to a superfluid phase \cite{Stoof99}, the analog of the
superconducting (BCS) phase transition in metals
\cite{Bardeen57}. Very low temperatures and strong attractive
interactions in a two-component Fermi gas are favorable conditions to
reach this superfluid state. The interactions in atomic gases at low
temperature are usually described by a single parameter, the $s$-wave
scattering length $a$. This quantity can be tuned near a Feshbach
resonance where the sign and magnitude of $a$ can be adjusted by means
of an external magnetic field $B$ \cite{Stoof93, Feshbach}. Therefore,
when entering the regime of strong interactions, dilute Fermi gases
offer unique opportunities to test new theoretical approaches. For
instance, near resonance, the critical temperature for superfluidity
has been predicted to be as high as 0.25-0.5 $T_\mathrm{F}$
\cite{Holland01}, a temperature range experimentally accessible.

\begin{figure}[htbp]
\begin{center}
\epsfig{file=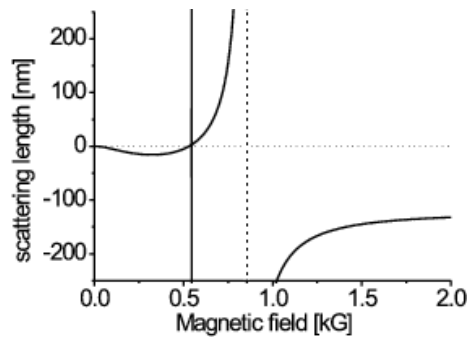, width=0.7\linewidth}
\caption{ \label{fig:Feshbach} Predicted scattering length $vs$ magnetic field in
$^6$Li $|F,m_F\rangle=|1/2,1/2\rangle$, $|1/2,-1/2\rangle$ mixture.}
\end{center}
\vspace{-0.8cm}
\end{figure}

Both $^{40}$K and $^6$Li possess Feshbach resonances at convenient
magnetic fields \cite{Houbiers97,Jin02}. In Fig.\,\ref{fig:Feshbach}
is plotted the theoretical scattering length for the mixture of the
two lowest spin states of $^6$Li, $|F,m_F\rangle=|1/2,1/2\rangle$,
$|1/2,-1/2\rangle$, calculated from updated potentials consistent with
the recently measured zero crossing of $a$ at $B \simeq 530\,$G
\cite{Jochim02,OHara02a}.
Beside the broad resonance near $855\,$G on which we concentrate in
this work, there is a very sharp resonance at $545\,$G, and increased atom losses have been measured peaking
around $545\,$G and $680\,$G \cite{Dieckmann02}. Recently, strong
interactions in this Fermi mixture have been demonstrated at a fixed
magnetic field of $930\,$G through anisotropic expansion of the gas
\cite{OHara02}.  In this Letter we study the atomic interactions
through the entire range of the Feshbach resonance, from $600\,$G up
to $1.3\,$kG. We measure three physical quantities\,: the interaction
energy with positive and negative values, the anisotropy of the atomic
cloud during the expansion, and the atom loss.  We find the resonance
at $800(40)\,$G, where the trapped gas is most strongly interacting,
with a ratio between the interaction energy and the kinetic energy
reaching $-0.3$.  Surprisingly, we observe a large shift ($\sim
80\,$G) between this position, and both the location of maximum loss
in the gas and the change of sign of the interaction energy. Our
results are in partial agreement with the physical picture of a
Feshbach resonance in the region of strong interactions but also
reveal unexpected effects. Finally they point towards the best
experimental conditions to search for the superfluid transition in $^6$Li.

Our experimental approach to measure the interaction energy of a two
component Fermi gas is based on the analysis of time of flight (TOF)
images of atoms released from an anisotropic trap. The energy of the
trapped gas, prepared at any value of the $B$ field, is the sum of
potential, kinetic and interaction energies,
$E_{pot}+E_{kin}+E_{int}$. Switching off abruptly the trapping
potential ($E_{pot}\rightarrow0$), the gas is released and expands for a
variable time before an absorption image is recorded.  At long
expansion time, the spatial distribution of the cloud reflects the
velocity distribution.  This procedure is done routinely for BEC
studies \cite{Fermi99} and has been recently investigated
theoretically for Fermi gases in \cite{Menotti02}. The novelty of our
approach is in what follows. Because of the low inductance of the
coils used to produce the magnetic field $B$, this field can be
switched off rapidly ($\leq 20\,\mu$s) at any desired time during the
expansion of the atomic cloud. For $B=0$ the atoms have negligible
interactions since $a\simeq 0$ (Fig.\,\ref{fig:Feshbach}). As a
consequence, the expansion of the gas can be recorded without atomic
interactions during the time of flight period ($B=0$), or with
interactions during the TOF period ($B\neq 0$). Since expansions with $B=0$
are ballistic, they reflect the kinetic energy of the initially
trapped gas, $E_{kin}$.  On the other hand, for TOF with $B\neq 0$,
the interaction energy is converted into kinetic energy during the
early stage of the expansion ($\lesssim 150 \, \mu$s). TOF images at
long time reflect the released energy, $E_{rel}=E_{kin}+E_{int}$
\cite{analyse}. Comparisons between TOF images with $B=0$ and $B\neq0$
allow us to simply deduce the ratio $E_{int}/E_{kin}$.  We operate at
temperatures between $0.5\,T_\mathrm{F}$ and $T_\mathrm{F}$, where
Fermi degeneracy does not play an important role. Even in this nearly
classical regime, the gas is found to be strongly interacting. The
observed expansions with $B\neq0$ are anisotropic because of
collisonal hydrodynamics, as predicted in \cite{Kagan97,Arimondo99},
and observed in \cite{Shvarchuck02,OHara02,Jinpreprint}.

Our experimental setup has been described previously
\cite{Khaykovich02,Schreck01}. A gas of $3 \times 10^5$ $^6$Li atoms
is prepared in the absolute ground state $|1/2,1/2\rangle$ in a Nd-YAG
crossed beam optical dipole trap at a temperature of $10\,\mu$K. The
horizontal beam (resp.\,vertical) propagates along $x$ ($y$),
has a maximum power of 2.5 W (2.5 W) and a waist of $\sim
25\,\mu$m ($\sim 30\,\mu$m). At full power the trap oscillation
frequencies are $\omega_x/2\pi=2.2(2)\,$kHz,
$\omega_y/2\pi=3.0(3)\,$kHz, and $\omega_z/2\pi=3.7(4)$\,kHz, as
measured by parametric excitation.  Using a radio frequency field, we
drive the Zeeman transition between $|1/2,1/2\rangle$ and
$|1/2,-1/2\rangle$ to prepare a balanced mixture of the two states at
any chosen value of $B$ between $2\,$G and $1.3\,$kG. Using a
Stern-Gerlach method we check that the populations in both states are
equal to within 10\%.

\begin{figure}[htbp]
\begin{center}
\epsfig{file=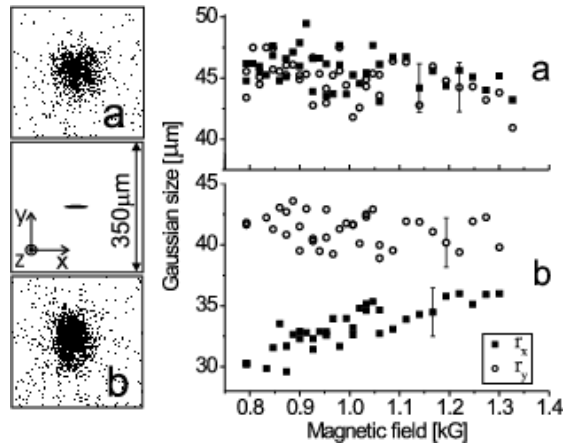, width=0.85\linewidth}
\caption{ \label{fig:detection}
\label{fig:sigmahighfield} Left: 
Geometry of the trapped atomic cloud (center) and expansion images
without magnetic field (a) ($B=0$), and with magnetic field (b) ($B\neq
0$). Right: Corresponding gaussian sizes of the expanded clouds along
x (squares) and y (open circles) $vs$ magnetic field. a : Time of
flight images after $650\,\mu\mathrm{s}$ expansion with $B=0$. b:
Images after $400\,\mu\mathrm{s}$ expansion with $B\neq 0$ and
$250\,\mu\mathrm{s}$ with $B=0$. Images shown on the left are for
$B=900\,$G.}
\end{center}
\vspace{-0.5cm}
\end{figure}

We observe that atom losses occur with very different rates below and
above resonance. Near $720\,$G the lifetime of the gas is on the order
of 10\,ms, whereas near $900\,$G, the lifetime is in excess of
10\,s. The latter is surprisingly large in comparison with similar
situations for bosons near a Feshbach resonance \cite{Inouye98}.
Therefore we performed two sets of experiments\,: one with the spin
mixture prepared above resonance ($1060\,$G), and the other below
resonance ($5\,$G). In the first set, evaporative cooling is performed
by lowering the power of the vertical beam. It produces $2N=7 \times
10^4$ atoms at $T\simeq 0.6\,T_{\mathrm F}$ with $k_\mathrm{B}
T_\mathrm{F}=\hbar\bar{\omega} \left(6 N\right)^{1/3}$ and
$\bar{\omega}=\left(\omega_x \omega_y \omega_z\right)^{1/3}$. The trap
is nearly cigar shaped with frequencies $\omega_x/2\pi=1.1\,$kHz,
$\omega_y/2\pi=3.0\,$kHz, and $\omega_z/2\pi=3.2\,$kHz. The magnetic
field is adiabatically ramped in 50\,ms to a final value where TOF
expansion images are taken. The effect of strong interactions is
illustrated in Fig.\,\ref{fig:detection}, where both typical images
and a plot of the expanded cloud gaussian sizes $r_x$ and $r_y$ are
displayed. In Fig.\,\ref{fig:detection}a, expansions with $B=0$
reveal the initial momentum distribution of the cloud which is
isotropic as expected. We deduce the total kinetic energy $E_{kin}$ of
the trapped gas mixture from gaussian fits to this distribution. For our 
moderate quantum degeneracy, the
temperature can be estimated from $k_\mathrm{B}T=2E_{kin}/3$.  We find
$T\simeq 3.5\, \mu$K$\simeq 0.6 \,T_{\mathrm F}$, constant for fields
between $0.8\,$kG and $1.3\,$kG.

In Fig.\,\ref{fig:detection}b, expansions with $B\neq 0$ are
anisotropic. Little expansion is seen along the weak axis of the
trap. The ellipticity of the cloud is inverted because of hydrodynamic
behavior during the expansion \cite{OHara02}. Collisions then
redistribute the gas energy in the direction of maximum density
gradient.  The anisotropy $r_y/r_x$ ranges from $1.1$ at large $B$
field to $1.4$ near $0.8\,$kG whereas the hydrodynamic scaling
equations \cite{Kagan97} predict an anisotropy of $1.53$ in the fully
hydrodynamic regime. The usual criterion for hydrodynamicity is found
from the ratio $R$ of the mean free path $\lambda_0=(n_0
\sigma)^{-1}$, where $n_0$ is the peak density, over the radial size
$r_{rad}$ of the cloud. For a classical gas neglecting mean field
interactions we find\,:
$$R=\frac{\lambda_0}{r_{rad}}=\frac{(2\pi)^{3/2}}{N
\sigma}\left(\frac{k_BT}{m\bar{\omega}^2}\right)\frac{\omega_{rad}}{\bar{\omega}}$$
$R \ll 1$ ($R \gg 1$) corresponds to the hydrodynamic (collisionless)
regime. With the predicted value of $a=-185\,$nm at $1060\,$kG, and
$\sigma=4\pi a^2$, $R=0.03$.  Therefore in the early stages of the
expansion, the gas is hydrodynamic as in \cite{Shvarchuck02,OHara02,
Jinpreprint}. Furthermore, with these large values of $a$ and our
typical temperatures, the scattering cross section is energy
dependent. At $B=1060\,$G, $|ka|=0.95$, where
$k=\sqrt{mk_BT/2\hbar^2}$ is the typical relative momentum of two
colliding atoms. The cross section is reduced and becomes unitarity
limited, $\sigma=4\pi a^2/(1+k^2a^2)\simeq {4\pi}/{k^2}$ for $|ka| \gg
1$ \cite{Combescot03}.  Consequently, as the magnetic field is
decreased below 1.3\,kG, the gas gradually enters deeper in the
unitarity regime.

This effect has an important consequence on the gas behavior during
hydrodynamic expansion: since the relative momentum $k$ of the colliding
atoms decreases as $k \propto n^{1/3}$ (where $n$ is the density)
\cite{liouville}, $\sigma$ increases.  For a spherical cloud,
the hydrodynamic factor $R$ would remain constant during expansion
while for a cigar shaped cloud, the expansion is mostly 2D, and $R$
decreases like $n^{1/6}$. The gas becomes more hydrodynamic as the
expansion proceeds until this model breaks down when relative momenta
$k$ become too small to remain in the unitarity limit \cite{Shlyapnikov02}.
Then $R$ increases as $n^{-2/3}$ for a spherical geometry, and as
$n^{-1/2}$ for a cigar.  Consequently, the larger $|a|$, the longer
the expansion remains hydrodynamic and the stronger the anisotropy
is. When the $B$ field is decreased in Fig.\,\ref{fig:sigmahighfield},
the anisotropy increase indicates that the scattering length becomes
more and more negative. Therefore, this puts an upper bound of $\simeq
800\,$G for the position of the Feshbach $s$-wave resonance.

\begin{figure}[htbp]
\begin{center}
\epsfig{file=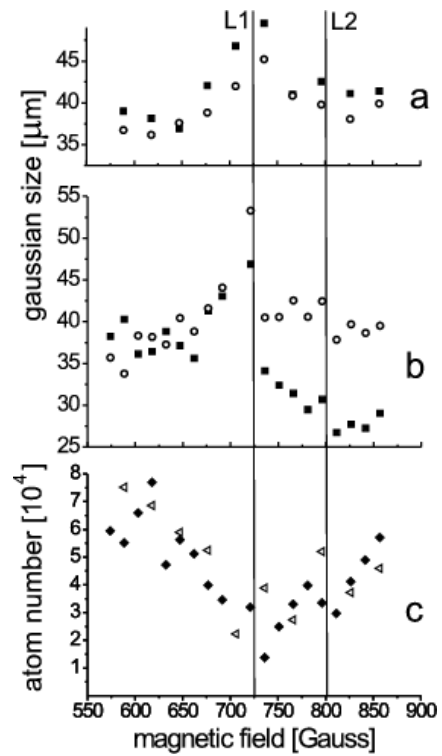, width=0.65\linewidth}
\caption{\label{fig:lowfield} a and b: Gaussian sizes along $x$ and
$y$ $vs$ magnetic field. The detection conditions and symbols are the
same as in Fig.\,\ref{fig:sigmahighfield}.  a: Expansion with
$B=0$. b: Expansion with $B \neq 0$ c: Number of detected atoms $vs$
magnetic field. Open triangles and black diamonds correspond to graph
a and b, respectively. Line L1 corresponds to the maximum of loss, and
L2 to the resonance position.}
\end{center}
\vspace{-0.8cm}
\end{figure}

In the second set of experiments, we focused on the lower magnetic
field values between 550 G and 820 G, the region where losses have
been observed.  The mixture is prepared at low magnetic field, and the
field is ramped up to 320 G, where the scattering length $vs$ $B$
has a local minimum of $a=-8\,$nm.  Evaporative cooling is performed
there, leading to $T=2.4\, \mu$K and to $\omega_x/2\pi=0.78\,$kHz,
$\omega_y/\pi=2.1\,$kHz, and $\omega_z/2\pi=2.25$\,kHz. Finally, the
$B$ field is ramped to 555 G in 50 ms where $a\simeq 0$, and then in
10 ms to different values between $600\,$G and $850\,$G where TOF
expansions are recorded.  The gaussian widths and number of detected
atoms $vs$ $B$ are plotted in Fig.\,\ref{fig:lowfield}.
Comparing with the data of Fig.\,\ref{fig:sigmahighfield}, we observe
several new features. First, the number of detected atoms has a
pronounced minimum near $725\,$G, a position compatible with
previously published results dealing with losses \cite{Dieckmann02}. These losses are associated with
a strong heating of the cloud (Fig.\,\ref{fig:lowfield}a).

Second, in figure \ref{fig:lowfield} (b) for $B\leq 675\,$ G, the
expansion is isotropic, consistent with a collisionless
regime. Here $a$ is predicted to be $0 \leq a \leq 55\,$nm. Since $ka
\leq 0.35$, the scattering cross section is independent of energy and
the gas is not hydrodynamic ($R \ge 1$).  Above $700 \,$ G, a
pronounced asymmetry in TOF images appears abruptly. The gas
enters the hydrodynamic regime (and unitarity limit) with the same
inversion of ellipticity as in Fig.\,\ref{fig:sigmahighfield}, a
signature of strong interactions.  We expect that on resonance, the
elastic cross section is unitarity limited for all relevant values of
energy. Therefore, the maximum anisotropy, which we measure at
$B=800(40)\,$G, locates the peak of the Feshbach resonance, in
agreement with the predicted position of 855(30) Gauss. A unique
feature of the resonance is the very large shift ($\simeq 80\,$G)
between the resonance peak and the maximum of loss and heating.  This
can be qualitatively explained by the creation for $a\ge 0$ of weakly
bound molecules by three-body recombination. According to
\cite{Petrov02}, in this process the binding energy of the molecule
$\hbar^2/ma^2$ for large $a$, is dissipated into kinetic energy of the
atom + molecule system. For very large values of $a$, the binding
energy is small and the heating associated with three-body
recombination is negligible, and the molecules remain trapped.  For
small or intermediate values of $a$, the collision products are highly
energetic, resulting in strong heating and loss. At the maximum of
loss ($720\,$G), $a=102\,$nm and $\hbar^2/k_\mathrm{B}ma^2=7\,\mu$K is
on the order of the trap depth in the weakest direction $x$.

\begin{figure}[htbp]
\begin{center}
\epsfig{file=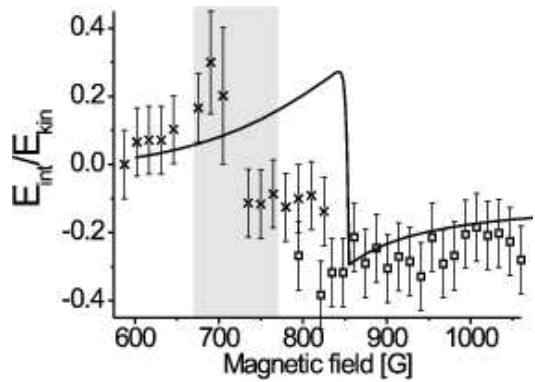, width=0.8\linewidth}
\caption{\label{fig:beta} Ratio of interaction energy over kinetic
energy $vs$ magnetic field. Open squares: atomic samples prepared
above resonance as in Fig.\,\ref{fig:sigmahighfield}.  Crosses:
average of three sets of data, recorded in conditions of
Fig.\,\ref{fig:lowfield}.  The difference between crosses and open
squares in the overlap region is due to different experimental
conditions described in the text. Solid line\,: Mean field theory
calculated with a total number of atoms of $7 \times 10^4$ and a
kinetic energy of $E_{kin}/k_B=5.25\,\mu$K, values corresponding to
the open squares conditions. The gray area indicates the region of losses.
}
\end{center}
\vspace{-0.8cm}
\end{figure}
As explained above, the interaction energy can be calculated directly
from the difference of release energies, obtained from the gaussian
sizes of Figs.\,\ref{fig:sigmahighfield} and \ref{fig:lowfield}. In
Fig.\,\ref{fig:beta} is plotted the ratio of interaction energy and
kinetic energy $vs$ magnetic field. As $B$ is increased above 550 G, $E_{int}$
first goes up, due to increasing $a$, in accordance with mean-field
theory.  Near $720\,$G, $E_{int}$ changes sign abruptly.  The ratio
$E_{int}/E_{kin}$ then exhibits a plateau near $-0.12$ up to
$850\,$G. Note that, for the data at low field
(Fig.\,\ref{fig:lowfield}), the number of atoms and the temperature
are not constant as a function of $B$.  From the data at higher
fields, we observe in Fig.\,\ref{fig:beta} that the interaction energy
becomes less and less negative above $800\,$G. The apparent
discrepancy between the two sets of data in the overlap region can be
explained by different experimental conditions and, in particular,
different atom numbers, temperature, and confinement. The line in
Fig.\,\ref{fig:beta} shows the result of a mean field calculation
without any fit parameters based on a full energy dependent scattering
phase shift. The mean field energy $E_{\rm MF}$ is then obtained from
a self-consistent distribution function that contains the self-energy
and trapping potential. The kinetic energy is calculated from the same
distribution function.  Theory and experiment agree quantitatively 
for $B\geq850\,$G where $a<0$, and for $B\leq 720$G, where
$a>0$. However, the observed shift between the predicted position of
the resonance and the change of sign of the mean field $720(20)\,$G is
not reproduced by this theory.  A possible explanation could be that
dimer molecules, which are expected to be present for $a>0$ and close
to resonance, are not part of the calculation.  A large negative
partial mean field contribution due to atom-molecule and
molecule-molecule interactions could shift the position of the
transition from positive to negative mean field.

In summary, we have studied a Fermi gas mixture in the strongly
interacting regime near a Feshbach resonance. The position of the
resonance is found in agreement with theoretical expectations. New
features have been observed which may be the signature of richer
physics, for instance molecule formation \cite{Wiemann02}.
Anisotropic expansions are observed both for repulsive and attractive
mean field interactions, in a moderately degenerate Fermi gas. This is
interpreted in terms of collisional hydrodynamics without invoking
Fermi superfluidity.  Prospects for producing deeply degenerate Fermi
mixtures in the superfluid state and for investigating the transition
between molecular condensates and superfluid Fermi gases are
promising.

We are grateful to L.\,Carr, Y.\,Castin, C.\,Cohen-Tannoudji,
R.\,Combescot, J.\,Dalibard, D.\,Gu\'ery-Odelin for useful
discussions. This work was supported by CNRS, Coll\`ege de France,
Region Ile de France, and EU (TMR network ERB FMRX-CT96- 0002).
Laboratoire Kastler Brossel is {\it Unit\'e de Recherche de l'Ecole
Normale Sup\'erieure et de l'Universit\'e Pierre et Marie Curie,
associ\'ee au CNRS}.


\begin{thebibliography}{99}
\bibitem{Fermi99} See for instance, Proc. of the Int. School of Physics "Enrico Fermi", M.
Inguscio, S.\,Stringari, and C.\,Wieman eds., It. Phys. Soc. (1999)
\bibitem{DeMarco99}
B.\,DeMarco and D.\,Jin, Science {\bf 285}, 1703 (1999).
\bibitem{Truscott01}
A.\,G.\,Truscott {\it et al.}, Science {\bf 291}, 2570 (2001).
\bibitem{Schreck01}
F.\,Schreck {\it et al.}, Phys. Rev. Lett., {\bf 87}, 080403 (2001).
\bibitem{Dieckmann02}
K.\,Dieckmann {\it et al.}, Phys. Rev. Lett {\bf
89}, 203201 (2002).
\bibitem{Modugno02}
G.\,Modugno {\it et al.}, Science {\bf 297},2240 (2002).
\bibitem{OHara02}
K.\,O'Hara {\it et al.}, Science {\bf 298}, 2179 (2002).
\bibitem{Stoof99}
A.\,G.\,Leggett, J. Phys. (Paris) {\bf C7}, 19 (1980).
\bibitem{Bardeen57}
J.\,Bardeen, L.\,Cooper and J.\,Schrieffer, Phys. Rev. A {\bf 108},
1175 (1957).
\bibitem{Feshbach}
H.\,Feshbach, Ann. Phys. (N.Y.) {\bf 5}, 357 (1958), {\bf 19}, 287 (1962).
\bibitem{Stoof93}
E.\,Tiesinga, B.\,J.\,Verhaar, H.\,T.\,C.\,Stoof Phys. Rev. A {\bf 47},
4114 (1993).
\bibitem{Holland01}
M.\,Holland, S.\,J.\,J.\,M.\,F.\,Kokkelmans, M.\,L.\,Chiofalo and R.\,Walser, Phys. Rev.
Lett., {\bf 87} 120406 (2001).
\bibitem{Houbiers97} M.\,Houbiers {\it et al.}, Phys. Rev. A {\bf 57},
R1497 (1998).
\bibitem{Jin02}
T.\,Loftus {\it et al.}, Phys. Rev. Lett. {\bf 88}, 173201 (2002).
\bibitem {OHara02a}
K.\,O'Hara {\it et al.}, Phys. Rev. A, {\bf 66}, 041401(R) (2002).
\bibitem{Jochim02}
S.\,Jochim {\it et al.}, Phys. Rev. lett., {\bf 89},273202 (2002).
\bibitem{Menotti02}
C.\,Menotti, P.\,Pedri, S.\,Stringari, Phys. Rev. Lett. {\bf 89}, 250402 (2002). 
\bibitem{analyse} We derive $E_{rel}$ as if the whole expansion had
been ballistic.  Simulations of our hydrodynamic expansions show that
this approximation induces a maximum error of 3\%.
\bibitem{Kagan97} Y.\,Kagan, E.\,L.\,Surkov, G.\,V.\,Shlyapnikov,
Phys. Rev. A, {\bf 55}, R18 (1997).
\bibitem{Arimondo99}
E.\,Arimondo,
E.\,Cerboneschi, H.\,Wu, 
p.\,573 in Ref.\,\cite{Fermi99}.
\bibitem{Jinpreprint} C.\,A.\,Regal, D.\,Jin, E-print: cond-mat/0202246.
\bibitem{Shvarchuck02} I.\,Shvarchuck {\it et al.},
Phys. Rev. Lett. {\bf 89}, 270404 (2002).
\bibitem{Khaykovich02} L.\,Khaykovich {\it et al.}, Science, {\bf 296},
1290-1293 (2002).
\bibitem{Inouye98} S.\,Inouye {\it et al.}, Nature {\bf 392}, 151
(1998).
\bibitem{liouville} In an adiabatic transformation of the gas, the
phase space density is constant during the time of flight.
\bibitem{Combescot03} In the strongly degenerate regime, this formula
should be modified. See R.\,Combescot, E-print: cond-mat/0302209.
\bibitem{Shlyapnikov02} G.\,V.\,Shlyapnikov, in proc. of the 18$^{th}$
Int. Conf. on At. Phys., H. R. Sadeghpour, D. E. Pritchard, and
E.\,J.\,Heller eds., World Scientific (2002).
\bibitem{Petrov02} 
D.\,S.\,Petrov, Phys. Rev. A {\bf 67}, 010703 (2003);
L.\,Pricoupenko, E-print: cond-mat/0006263.
\bibitem{Wiemann02}
E.\,A.\,Donley {\it et al.}, Nature {\bf 417}, 529 (2002).



\end{thebibliography}
\end{document}